\documentstyle[aasms4,12pt]{article}
\begin{document}

\title{STARBURST-DRIVEN STARBURSTS 
       IN THE HEART OF ULTRALUMINOUS INFRARED GALAXIES}

\author{Yoshiaki Taniguchi$^1$, Neil Trentham$^2$, and Yasuhiro Shioya$^1$}

\vspace {1cm}

\affil{$^1$Astronomical Institute, Tohoku University, Aoba, Sendai 980-8578, Japan}
\affil{$^2$Institute of Astronomy, University of Cambridge, Madingley Road, Cambridge
       CB3 0EZ, UK}


\begin{abstract}

There is increasing evidence for the presence of
blue super star clusters in the central regions 
of ultraluminous infrared galaxies like Arp 220.
Ultraluminous galaxies are thought to be triggered by galaxy
mergers, and it has often been argued 
that these super star clusters may form
during violent collisions between gas clouds
in the final phase of the mergers.
We now investigate another set of models 
which differ from previous ones in that the formation
of the super star clusters is linked directly to the very intense
starburst occurring 
at the very center of the galaxy.
Firstly we show that a scenario in which the super star clusters
form in material compressed by shock waves originating from the
central starburst is implausible because the objects so produced
are much smaller than the observed star clusters in Arp 220.
We then investigate a scenario (based on the Shlosman-Noguchi
model) 
in which the infalling dense gas disk
is unstable gravitationally and collapses to form massive 
gaseous clumps.
Since these clumps are exposed to the 
external high pressure driven by the superwind
(a blast wave driven by a collective effect of a large number of supernovae
in the very core of the galaxy),
they can collapse and then massive star formation may be induced in them.
The objects produced in this kind of collapse have properties consistent 
with those of the observed super star clusters in the center of Arp 220.
\end{abstract}



\section{INTRODUCTION}

Galaxy mergers provide a mechanism for
the metamorphosis of galaxies and sometimes
trigger intense star formation events (starbursts) 
in the central regions of the 
merger remnants (Toomre 1977; Schweizer 1982). 
When such intense starbursts happen, 
ultraluminous infrared galaxies (ULIGs) result 
because of the enormous energy released from 
the central regions of the galaxies
(Soifer et al. 1984; Wright et al. 1984; Sanders et al. 1988;
Sanders \& Mirabel 1996). 
One of the remarkable properties found in such 
galaxy mergers is that a number of blue super star clusters (SSCs) are resided in their
central regions (Lutz 1991; Ashman \& Zepf 1992; Holtzman et al. 1992; 
Shaya et al. 1994; see for a review Zepf \& Ashman 1993).
One plausible idea is that
such SSCs may be formed
by violent collisions among gas clouds
in the final phase of the galaxy mergers 
(Ashman \& Zepf 1992; Zepf \& Ashman 1993).

The SSCs in the mergers
are then distributed over the circumnuclear 
regions which are spatially close to  
the currently active starburst core (cf. Lutz 1991; Shaya et al. 1994).
Such SSCs, while common in the cores of ULIGs, are rarely seen in galaxies
where there is no evidence for an intense burst of star formation at the
very center.
Therefore, it is intriguing to investigate physical relationships between the
circumnuclear SSCs and the nuclear starburst
and address the question: is the formation of
SSCs attributed to some effects of the nuclear starburst?
This is the subject of the present paper. 
The link between the concentrated starburst in the core and the formation
of the SSCs is a fundamental difference between the models we discuss and
those that rely on violent collisions mentioned in the previous
paragraph. 

The nearest and best-studied ULIG is Arp 220 ($z = 0.02$) and we use it
as a basis for comparison in the present work.
In this galaxy, two nuclei exist with a 
projected separation of 350 pc (e.g., Condon et al. 1991).
Recent optical and near-infrared imaging made with the Hubble Space Telescope
has shown that about a dozen of star forming clumps are distributed in the
circumnuclear region with a radius of several hundred pc in Arp 220
(Shaya et al. 1994; Scoville et al. 1998). In particular, the western nucleus
of Arp 220 is surrounded by several clumps which are distributed along the
north-south direction.
In Figure 1 we show a schematic illustration of
the spatial distribution of SSCs in Arp 220.
These clusters are distributed within a radius of 
700 pc from this western nucleus.
The masses of SSCs are estimated to be $\sim 10^6$ - $10^8 M_\odot$ 
although these numbers depend on the assumed 
stellar populations (Shaya et al. 1994).

These SSCs are more massive and larger ($\sim$ 10 pc) 
than typical globular clusters in the Galaxy. 
Nevertheless, it is possible
that the formation of SSCs provides important hints 
on the formation of globular clusters.
Recent theoretical studies have suggested that globular clusters are
envisaged as forming in giant molecular clouds, in the same way as
we see star-cluster formation in our Galactic disk (Harris \& Pudritz 1994;
McLaughlin \& Pudritz 1996; Elmegreen \& Efremov 1997). 
Early in the formation of the Milky Way, which is when the globular
cluster formed, there may also have been considerable concentrated
star formation, as the bulge and the central part of the disk formed.
This hypothetical comparison provides a secondary motivation for
investigating the possible 
role of these concentrated starbursts in galaxy cores
in forming massive star clusters.

\section{STARBURST-DRIVEN STARBURSTS IN ARP 220}

Arp 220 shows a double-bubble structure of extent 13 kpc
(Heckman et al. 1987, 1996 see Figure 1).  Such a morphology is 
evidence for a superwind i.e.~a 
the blast wave driven by a large number of supernovae
(Tomisaka \& Ikeuchi 1988; 
Heckman, Armus, \& Miley 1990; Suchkov et al. 1994).
A typical galactic-scale nuclear
starburst forms numerous ($> 10^4$) massive stars 
during a short period ($\sim 10^7$ years;
Weedman et al. 1981, Balzano 1983). 
Therefore, a burst of supernovae occurs 
$\sim 10^7$ years after the onset of the starburst.
Since these numerous supernovae release a huge amount of kinetic
energy into the circumnuclear gas,
the circumnuclear gas is thermalized 
and then ejected from the starburst region
as a superwind.  This superwind interacts with ambient gaseous matter,
leading to the formation of shocked gaseous shell in the circumnuclear region. 
If there is a dense gaseous disk surrounding the nuclear starburst regions,
this disk can collimate the superwind 
along the direction perpendicular to the disk (Tomisaka \& Ikeuchi 1988);
such a disk is observed in Arp 220 by Scoville et al.~(1997).
Gas in the inner parts of this dense
collimating disk is will be affected by both the 
dynamical (e.g., shocks)
and thermal (e.g., compression by the high-pressure hot gas) effects of 
the superwind and it is these processes that we consider in the
rest of this paper.  The physical processes we describe here are quite
distinct from what might be happening in the outer regions of the galaxy
where the superwind interacts with the ambient hot diffuse coronal 
gas.  At this stage we note that in Figure 1,
the SSCs are distributed along the 
direction almost perpendicular to the double-bubble structures;
this is what we expect if the SSCs form in the dense collimating gas disk.

The inner parts of the dense gas disk should experience a significant 
dynamical effect from the superwind.
While direct evidence of shocked gas in Arp 220 is not observed,
we note that shocked regions are indeed observed in the nearer but
less luminous nuclear starburst galaxy 
M82 (Lester et al. 1990).
Although the actual gaseous matter in the nuclear region of Arp 220 is
almost certainly clumpy, it is instructive to derive order-of-magnitude
estimates of the dynamical effects of the superwind by considering
the effects of a superwind propagating through a homogenous medium.
The supernovae responsible for shocking the gas occur continuously
over a timescale longer than or comparable with the dynamical timescale
of the initial gas cloud, and
the evolution of the shocked material can be described by a 
superbubble model
(McCray \& Snow 1979; Koo \& McKee 1992a, 1992b;
Heckman et al. 1996; Shull 1995 and references therein).
The radius and velocity of the shocked shells formed in the dense gas 
disk at time $t$ are then

\begin{equation}
r_{\rm shell} \sim 440 L_{\rm mech, 43}^{1/5} n_{\rm H, 4}^{-1/5} t_7^{3/5} ~ {\rm pc},
\end{equation}

\noindent and

\begin{equation}
v_{\rm shell} \sim 26 L_{\rm mech, 43}^{1/5} n_{\rm H, 4}^{-1/5} t_7^{-2/5} ~ {\rm km~ s}^{-1},
\end{equation}

\noindent where $L_{\rm mech, 43}$ is the mechanical luminosity
released collectively from the supernovae  in units of 10$^{43}$ erg s$^{-1}$,
$n_{\rm H, 4}$ is the average hydrogen number density of the ISM in the dense gas disk,
assumed uniform, in units of $10^4$ cm$^{-3}$, and $t_7$ is the elapsed time
since the onset of the superwind in units of $10^7$ years.
For Arp 220, recent observations suggest that
$L_{\rm mech, 43} = 1$ (Heckman et al. 1996) and
$n_{\rm H, 4} = 1$ (Scoville et al. 1991, 1997).
Although we do not know physical conditions of nuclear gas disk
in the recent past 
(i.e., $\sim 10^7$ years ago), they may not be dramatically
different from those observed currently. 
We therefore assume  $n_{\rm H} = 10^4$ cm$^{-3}$.
In any case, the dependencies on $n_{\rm H}$ of both $r_{\rm shell}$ and 
$v_{\rm shell}$ are very weak (i.e., $\propto n_{\rm H}^{-1/5}$). 
Hence if our assumed values of $n_{\rm H}$ is incorrect by a factor
of two or three, the effects on the shell properties are very weak.
In order to explain the spatial extension of the double-bubble structures,
Heckman et al. (1996) estimated an  age of the superwind,
$t_7 \simeq 3$ given a low-density ambient gas ($n_{\rm H}
\sim 10^{-2}$ cm$^{-3}$). 
Using this age, we obtain $r_{\rm shell} \simeq 800$ pc for the dense gas disk.
This is approximately the extent of the SSCs.

Next we investigate the formation mechanism of the SSCs 
in the dense gas disk around the western nucleus.
Gaseous fragments can grow gravitationally in the shocked shell
formed by the dynamical effect of the superwind
(cf. Whitworth et al. 1994). 
Fragments that form within the shells experience a net inward
acceleration due to self-gravity and a net outward acceleration due to
the internal pressure.  Whitworth et al.~(1994) investigate the
balance between these two accelerations and show that the time-scale
for the growth of the fastest-growing fragments is
$t_{\rm fastest} \approx (2 c_{\rm s})/(G \Sigma)$
where $c_{\rm s}$ is the sound speed in the shell, and
$\Sigma$ is its surface mass density.
Non-linear fragmentation in the shell then happens first at a time
$t = t_{\rm fastest}$.  Noting that the surface mass density
$\Sigma = C {{n}_{\rm H} } m_{\rm H} r_{\rm shell}$,
where $C$ is a constant determined by the geometry ($C = 1/3$ for
a sphere) and $m_{\rm H}$ is the hydrogen atom mass,
we then find from equations (1) and (2) that fragmentation
within the shell first happens at a time
$t_{\rm f} \sim 5.7 \times 10^5 C_{0.33}^{-1/2} n_{\rm H, 4}^{-1/2} 
{\cal M}_{\rm c, 10}^{-1/2} ~ {\rm yr}$ at a radius
$r_{\rm f} \sim 80 L_{\rm mech, 43}^{1/5} C_{0.33}^{-3/10} n_{\rm H, 4}^{-1/2}
{\cal M}_{\rm f, 10}^{-3/10} ~ {\rm pc}$.
Here ${\cal M}_{\rm f, 10}$ is the Mach number when these
fragments first appear in units of 10, equal to $v_{\rm shell}/c_{\rm s}$
(here it is assumed that ${\cal M}_{\rm f} \gg 1$; Whitworth
et al.~1994).
Estimating $c_{\rm s}$ is difficult because the turbulent pressure in
the shell is much greater than the thermal pressure. From equation (2),
the shell is moving outward at a velocity $v_{\rm f}$ when
the fragments first appear, where
$v_{\rm f} \sim 81
L_{\rm mech, 43}^{1/5}
{\cal M}_{\rm f, 10}^{1/5} {\rm km~s}^{-1}$.
Thus $v_{\rm f}$ is almost independent of $c_{\rm s}$, so that our lack of
knowledge of the sound speed is unimportant in determining the
velocity of the shell when the fragments form.
Following Whitworth et al.~(1994), we can estimate the mass and size of
the fragments:
$M_{\rm f} \sim c_{\rm s}^{7/2} (G^3 n_{\rm H} m_{\rm H} v_{\rm f})^{-1/2}
\sim  7.9 \times 10^4
c_{\rm s, 10}^{7/2}
n_{\rm H, 4}^{-1/2}
v_{\rm f, 81}^{-1/2} ~ M_\odot$ and
$l_{\rm f} \sim c_{\rm s}^{3/2} (G n_{\rm H} m_{\rm H} v_{\rm f})^{-1/2}
 \sim  5.4 c_{\rm s, 10}^{3/2}
n_{\rm H, 4}^{-1/2}
v_{\rm f, 81}^{-1/2}$ pc
where $c_{\rm s}$ and $v_{\rm f}$ are expressed in units of 
10 km s$^{-1}$ and 81 km s$^{-1}$, respectively. 
We note that the local sound speed in the molecular gas clouds
is $c_{\rm s}^{\rm local} \sim 0.5 T_{50}^{1/2}$ km s$^{-1}$
where $T_{50}$ is the temperature of molecular gas in units of 50 K.
However, when we discuss the fragmentation in the shell, 
it is more reasonable to regard the random velocity of clouds as an effective
sound speed -- this bulk velocity is much larger 
than the local thermal velocity (e.g., 90 km s$^{-1}$ in Arp 220;
Scoville et al. 1998).
The conclusion from this simple calculation is therefore: 
fragments of gas that grow gravitationally in the shocked shell
are too small ($\sim 10^4 M_\odot$ and $\sim$ 5 pc)
to be the SSCs
observed in Arp 220.
Adopting a smaller sound speed (which is the main uncertainty in the
above calculation) makes this statement even stronger.

Therefore we have to look for another formation mechanism of the
progenitor of SSCs.  We consider the instability of 
the dense gas disk in the central region of Arp 220 due to  
its self gravity.  We draw on the results of Shlosman \& Noguchi, who
investigated the general problem of the gravitational instability of nuclear
gas disks in detail.
They estimated the mass of a superclouds forming in such a disk as 

\begin{equation}
M_{\rm scl} = \pi \left( \frac{\lambda_{\rm crit}}{2} \right)^2 \Sigma_{\rm g}
\end{equation}

\noindent where  $\Sigma_{\rm g}$ is the surface mass density of the gas disk and 
$\lambda_{\rm crit}$ is a typical scale length of the supercloud given by

\begin{equation}
\lambda_{\rm crit} = \frac{2 \pi^2 G \Sigma_{\rm g}}{\kappa^2},
\end{equation}

\noindent where $\kappa$ is the epicyclic frequency at the region where
the gravitational instability occurs.  Equations (3) and (4) can be
expressed: 

\begin{equation}
M_{\rm scl}  \sim  
5.7 \times 10^9 \Sigma_{\rm g, 4}^{3} \kappa_3^{-4} \; \; M_{\odot},
\end{equation}

\noindent and 

\begin{equation}
\lambda_{\rm crit} \sim 
8.5 \times 10^2 \Sigma_{\rm g, 4} \kappa_3^{-2} \; {\rm pc},
\end{equation}

\noindent where $\Sigma_{\rm g, 4}$ is in units of $10^4 M_\odot$ pc$^{-2}$
and $\kappa_3$ is in units of $10^3$ km s$^{-1}$ kpc$^{-1}$.
Using the observed 
values of $\Sigma_{\rm g}$ and $\kappa$ (Scoville et al. 1998),
we obtain masses of the superclouds up to $\sim 10^9 M_\odot$ (see Table 1). 
These are the masses of the gaseous clouds which will only equal the
mass of forming stars, if the star formation efficiency is 1.
Adopting a more reasonable efficiency of 0.1 from observations of the
Galaxy, we derive a final stellar mass of 
$\sim 10^8 M_\odot$, which is
consistent with observation (Shaya et al. 1994).
As shown in Table 1, the masses of the gaseous clouds are  
less massive with increasing with radius.  On going from 0.1 kpc
to 1 kpc the clusters are approximately an order of magnitude
smaller assuming the same star formation efficiency everywhere.
Despite the (large) uncertainty due to extinction, this trend is
close to what is seen by Shaya et al.~(1994) -- see their Tables 1
and 3.
  
Finally, we consider how the massive and large proto-SSC clouds that
came from our stability analysis
evolve into the more compact and less 
massive SSCs seen in the HST images.
Once the superclouds are formed, they will collapse because of the
effect of external pressure (Kimura \& Tosa 1993).
Given the hydrogen number density and the kinetic temperature of the
superclouds, $n_{\rm H} \sim 10^4$ cm$^{-3}$ and 
$T_{\rm kin} \sim$ 10 K, 
the pressure supported by the random motion of gas is 
$P_{\rm int} \sim 2 \times 10^{-11}$ dyn cm$^{-2}$. 
Next we estimate the pressure of inter-supercloud gas.
Since the hot gas traced by soft X-ray emission is indeed 
associated with the nuclear region of Arp 220, it seems reasonable 
to assume $T_{\rm kin} \sim 10^7$ K (Heckman et al. 1996). 
However, there is no direct information on the inter-supercloud gas 
density $n_{\rm H,IC}$. Although there is considerable amounts
of gas ($\sim 10^{10} M_\odot$) 
in the nuclear region, most of this gas occurs in dense molecular  clouds
(Scoville et al. 1997). 
The only statement that we can make with any confidence about the
density of intercloud gas is that it must be intermediate between
the cloud density ($\sim 10^4 {\rm cm}^{-3}$) and the density of the
ambient coronal gas ($\sim 10^{-2}$ cm$^{-3}$; Heckman et al.~1996).   
A reasonable value is probably around 1 cm$^{-3}$, which is the
typical gas density of the cold ISM in our galaxy (Spitzer 1978). 
The external pressure provided by the hot gas is then
$P_{\rm ext} \sim 2 \times 10^{-9} (n_{\rm H,IC} / 1 {\rm cm}^{-3})$ 
dyn cm$^{-2}$ for $T_{\rm kin} \sim 10^7$K.
Therefore, the external pressure caused by the hot gas is 
higher than the internal one for any value of the intercloud
density higher than that of the coronal gas, and is much higher (by a
factor of 100) if $n_{\rm H,IC} \sim 1 {\rm cm}^{-3}$.
Since the cooling timescale of neutral gas in the fragments,
$t_{\rm cool} \sim 100 (n_{\rm H}/10^4 ~{\rm cm}^{-3})^{-1}$ yr
(Spitzer 1978), is quite short, stars could form just after the fragmentation. 
The stars may form through the process of gravitational instability
occurring inside the fragment. A typical length of the gravitational
fluctuation (the Jeans' length: Jeans 1929) is estimated to be 
$\lambda_{\rm J} \sim (\pi c_{\rm s}^{\rm local} / G \rho_{\rm gas})^{1/2}
\sim$ 1 pc where $\rho_{\rm gas} = n_{\rm H} m_{\rm H}$. 
Therefore, since a typical mass of the fluctuation 
amounts to $m \sim \lambda_{\rm J}^3 \rho_{\rm gas} \sim 100 M_\odot$,
it is likely that the fluctuation will evolve into a massive star
with a mass of $\sim 10 M_\odot$.
The lifetime of these massive stars is $\sim 10^7$ years.
The single fragment (comprising these stars)
experiences dynamical relaxation
on a time scale of 
$2\times 10^5 \lambda_{\rm crit, 100}^{3/2} M_{\rm scl, 8}^{-1/2}$ years
where $\lambda_{\rm crit, 100}$ is 
the diameter of the supercloud in units of 100 pc
and $M_{\rm scl, 8}$ is the mass of the supercloud in units of $10^8 M_\odot$.
These predicted values of the cloud size and mass are close to the
observed ones (several 10 pc and $10^6$ - $10^8 M_{\odot}$; Shaya et al. 1994)
in Arp 220.

Therefore adopting a Shlosman-Noguchi model to investigate the gravitational
instability of the circumnuclear disk in Arp 220
allows us to  explain the formation of SSCs found 
there.  The link between the formation of the clusters and the central
starburst that was discussed in Section 1 is clear -- the supernovae
occurring in the center lead to the formation of the superbubble, which in
turn provides an external pressure large enough to initiate collapse in
the gas disk. 

\vspace {0.5cm}

We would like to thank Makoto Tosa and an anonymous referee for useful
comments and suggestions.
This work was financially supported in part by Grant-in-Aids for the Scientific
Research (Nos.\ 07044054 and 10304013) of the Japanese Ministry of
Education, Science, Sports and Culture.
NT thanks the PPARC for financial support.
\newpage


\newpage


\begin{table}
\caption{Physical properties of the superclouds formed in the dense gas disk}
\begin{tabular}{ccccc}
\tableline
\tableline
$R$ & $\Sigma_{\rm g}$ & $\kappa$ & 
$\lambda_{\rm crit}$ & $M_{\rm scl}$ \\
(kpc) & ($M_{\odot}\; {\rm pc^{-2}}$) & (${\rm km \; s^{-1} \; {kpc}^{-1}}$) & 
(pc) & ($M_{\odot}$)\\
\hline
0.1 & $3.5 \times 10^4$ & $3.6 \times 10^3$ & 230 & $1.45 \times 10^9$ \\
0.3 & $0.8 \times 10^4$ & $1.5 \times 10^3$ & 302 & $5.76 \times 10^8$ \\
0.5 & $0.4 \times 10^4$ & $1.0 \times 10^3$ & 340 & $3.65 \times 10^8$ \\
1.0 & $0.1 \times 10^4$ & $0.5 \times 10^3$ & 340 & $9.12 \times 10^7$ \\
\tableline
\end{tabular}
\end{table}

\newpage


\centerline {\bf Figure captions}\par

\vspace{1cm}

\noindent {\bf Fig. 1:} The blue stellar clusters found in Shaya et al. (1994)
are overlaid on the 4.83 GHz radio continuum map obtained by Baan and 
Haschick (1995). The blue clusters are mostly distributed along the NS 
direction through the western radio nucleus.
A pair of the large-scale emission-line bubbles, traced by H$\alpha$+[NII]
$\lambda$6548, 6584 emission, is shown in the lower panel
taken from Heckman et al. (1996). The bubbles are roughly
elongated along the EW direction which is perpendicular to that of the blue
clusters.


\begin{references}
\reference{1}{Ashman, K. M., \& Zepf, S. E. 1992, ApJ, 384, 50}
\reference{1}{Baan, W. A., \& Haschick, A. D. 1995, ApJ, 454, 745}
\reference{1}{Balzano, V. A. 1983, ApJ, 268, 602}
\reference{1}{Condon, J. J., Huang, Z. -P., Yin, Q. F., \& Thuan, T. X. 1991, ApJ, 378, 65}
\reference{1}{Elmegreen, B. G., \& Efremov, Y. N. 1997, ApJ, 480, 235}
\reference{1}{Harris, W. E., \& Pudritz, R. E. 1994, ApJ, 429, 177}
\reference{1}{Heckman, T. M., Armus, L., \& Miley, G. K. 1987, AJ, 93, 276}
\reference{1}{Heckman, T. M., Armus, L., \& Miley, G. K. 1990, ApJS, 74, 833}
\reference{1}{Heckman, T. M., Dahlem, M., Eales, S. A., Fabbiano, G., \& Weaver, K.
              1996, ApJ, 457, 616}
\reference{1}{Holtzman, J. A., et al. 1992, AJ, 103, 691}
\reference{1}{Jeans, J. H. 1929, Astronomy and Cosmogony, 2nd ed. Cambridge, Eng.:
             (Cambridge University Press)}
\reference{1}{Kimura, T., \& Tosa, M. 1993, ApJ, 406, 512}
\reference{1}{Koo, B. -C., \& McKee, C. F. 1992a, ApJ, 388, 93}
\reference{1}{Koo, B. -C., \& McKee, C. F. 1992b, ApJ, 388, 103}
\reference{1}{Lutz, D. 1991, A \& A, 245, 31}
\reference{1}{McLaughlin, D. E., \& Pudritz, R. E. 1996, ApJ, 457, 578}
\reference{1}{McCray, R., \& Snow, T. P. 1979, ARA \& A, 17, 213}
\reference{1}{Sanders, D. B., et al. 1988, ApJ, 325, 74}
\reference{1}{Sanders, D. B., \& Mirabel, I. F. 1996, ARA \& A, 34, 749}
\reference{1}{Schweizer, 1982, ApJ, 252, 455}
\reference{1}{Scoville, N. Z. et al. 1998, ApJ, 492, L107}
\reference{1}{Scoville, N. Z., Sargent, A. I., Sanders, D. B., \& Soifer, B. T.
             1991, ApJ, 366, L5}
\reference{1}{Scoville, N. Z., Yun, M. S., \& Bryant, P. M. 1997, ApJ, 484, 702}
\reference{1}{Shaya, E. J., Dowling, D. M., Currie, D. G., Faber, S. M.,
              \& Groth, E. J. 1994, AJ, 1675}
\reference{1}{Shlosman, I., \& Noguchi, M. 1993, ApJ, 414, 474}
\reference{1}{Shull, M. J. 1995, in
              Airborne Astronomy Symposium on the Galactic
              Ecosystem, ASP. Conf. Ser. 73, 365}
\reference{1}{Soifer, B. T., et al. 1984, ApJ, 283, L1}
\reference{1}{Spitzer, L. Jr. 1978, Physical Processes in the Interstellar
              Medium, Chap. 6, 131}
\reference{1}{Suchkov, A. A., Balsara, D. S., Heckman, T. M., \& Leitherer, C.
              1994, ApJ, 430, 511}
\reference{1}{Tomisaka, K., \& Ikeuchi, S. 1988, ApJ, 330, 695}
\reference{1}{Toomre, A. 1977, in The Evolution of Galaxies
              and Stellar Populations, edited by B. M. Tinsley and R. B. Larson
              (Yale University Observatory) 401}
\reference{1}{Weedman, D. W., Feldman, F. R., Balzano, V. A., Ramsey, L. W.,
              Sramek, R. A., \& Wu, C. -C. 1981, ApJ, 248, 105}
\reference{1}{Whitworth, A. P., Bhattal, A. S., Chapman, S. J., Disney, M. J.,
              \& Turner, J. A. 1994, A \& A, 290, 421}
\reference{1}{Wright, G. S., Joseph, R. D., \& Meikle, W. P. S. 1984, Nature, 309, 31}
\reference{1}{Zepf, S. E., \& Ashman, K. M. 1993, MNRAS, 264, 611}
\end{references}
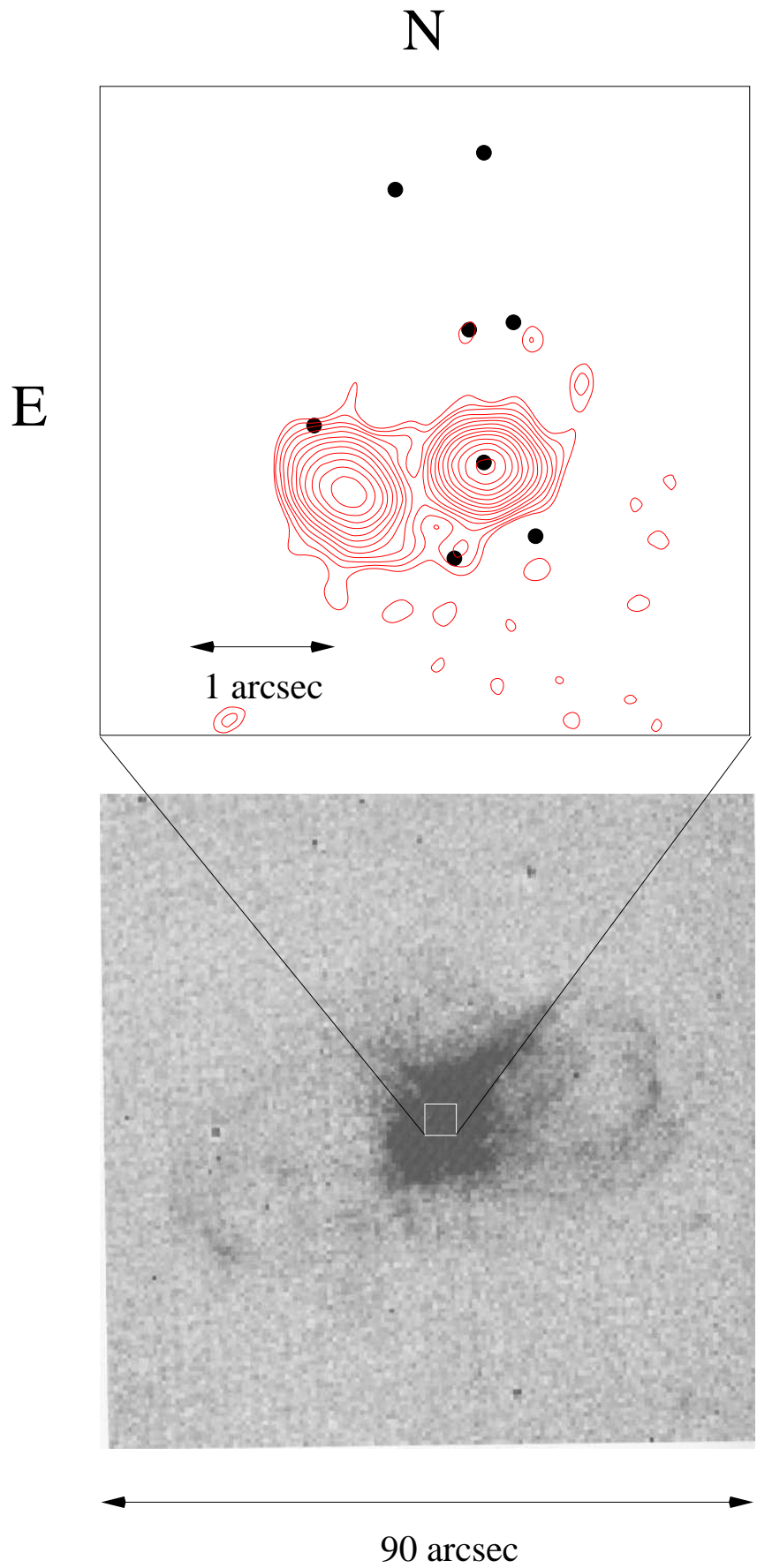
\end{document}